\documentclass[draft,wrr]{agutex}
 \usepackage{graphicx,tabularx}
 \graphicspath{ {./graphics/} }
 \DeclareGraphicsExtensions{.pdf,.jpg}
  \setkeys{Gin}{draft=false}
\authorrunninghead{TSE ET AL.}

\titlerunninghead{WIND-DRIVEN WETLAND WATER MOTIONS}

\begin{document}

%% ------------------------------------------------------------------------ %%
%
%  TITLE
%
%% ------------------------------------------------------------------------ %%

\title{Wind--driven water motions in wetlands with emergent vegetation}
%%evcp
%
% e.g., \title{Terrestrial ring current:
% Origin, formation, and decay $\alpha\beta\Gamma\Delta$}
%

%% ------------------------------------------------------------------------ %%
%
%  AUTHORS AND AFFILIATIONS
%
%% ------------------------------------------------------------------------ %%

\authors{Ian C. Tse, \altaffilmark{1} 
Cristina M. Poindexter, \altaffilmark{1,2},
and Evan A. Variano \altaffilmark{1}}

\altaffiltext{1}{Department of Civil and Environmental Engineering,
University of California, Berkeley, Berkeley, California, USA.}

\altaffiltext{2}{Department of Civil and Environmental Engineering,
California State University, Sacramento, Sacramento, California, USA.}

%% ------------------------------------------------------------------------ %%
%
%  ABSTRACT
%
%% ------------------------------------------------------------------------ %%

% >> Do NOT include any \begin...\end commands within
% >> the body of the abstract.

\begin{abstract}
Wetland biogeochemical transformations are affected by flow and mixing in wetland surface water. We investigate the influence of wind on wetland water flow by simultaneously measuring wind and surface water velocities in an enclosed freshwater wetland during one day of strong-wind conditions.
Water velocities are measured using a Volumetric Particle Imager while wind velocities are measured via sonic-anemometer. Our measurements indicate that the wind interacting with the vegetation canopy generates coherent billows and that these billows are the dominant source of momentum into the wetland water column. Spectral analysis of velocity timeseries shows that the spectral peak in water velocity is aligned with the spectral peak of in-canopy wind velocity, and that this peak corresponds with the Kelvin-Helmholtz billow frequency predicted by mixing layer theory. We also observe a strong correlation in the temporal pattern of velocity variance in the air and water, with high variance events having similar timing and duration both above and below the air-water interface. Water-side variance appears coupled with air-side variance at least down to 5 cm, while the theoretical Stokes' solution predicts momentum transfer down to only 2 mm assuming transfer via molecular viscosity alone. This suggests that the wind-driven flow contributed to significant mixing in the wetland water column.
\end{abstract}

%% ------------------------------------------------------------------------ %%
%
%  BEGIN ARTICLE
%
%% ------------------------------------------------------------------------ %%

% The body of the article must start with a \begin{article} command
%
% \end{article} must follow the references section, before the figures
%  and tables.

\begin{article}

%% ------------------------------------------------------------------------ %%
%
%  TEXT
%
%% ------------------------------------------------------------------------ %%

\section{Introduction}
Wetlands are hotspots of biogeochemical transformation, and these transformations can be influenced by the flow of wetland surface waters. For example, flow in wetlands affects the transport of pollen, seeds, and larva 
\citep{Middleton:2000bt,Ackerman:2002ck,Finelli:2000tm,Fonseca:1996hs}.
Gas transfer at the air--water interface is sensitive to surface flow, and influences both methane and mercury dynamics \citep{Poindexter:2013kg,Selvendiran:2008wp}.
%evcp
Several wetland-specific transport models have been developed to account for the unique hydrodynamics (including finer features like wake production) brought about by the presence of aquatic vegetation \citep{WRCR:WRCR8064,White:2003fn,Lightbody:2006vm,WRCR:WRCR11455}. These models describe transport in steady, unidirectional flows driven by gravity or barotropic pressure gradients, e.g., riverine or tidal forcing. There are some wetlands, however, that are isolated from such influences. In those, we expect exchanges with the atmosphere to account for a larger portion of the kinetic energy balance. %evcp
One important process that can drive surface-water flow in wetlands is thermal convection \citep{Nepf:1997uo,Jamali:2008cu,Poindexter:2013kg}.  

Wind action on uninterrupted water surfaces has been studied to understand waves \citep{Phillips}, coherent structures \citep{Shu} and air-water gas exchange \citep{Chu}.  How these processes change in the presence of vegetation has received little attention to date.  Only \citep{Bane} has discussed this issue, describing (among other things) the interacting dynamics of waves, turbulence, and water-depth.
%There is also evidence that strong winds can drive momentum exchange across the air-water interface, despite the sheltering effects of wetland vegetation. %evcp

Here, we investigate the influence of wind on hydrodynamics in wetland field site that is isolated from fluvial and tidal influence.  Previous field studies have shown evidence that wind can penetrate through a canopy to affect wetland surface water: in the Florida Everglades wind was found to correlate with suspended sediment concentrations in sloughs \citep{Noe:2009kx}; in a Massachusetts salt marsh, where wind caused velocity profiles to deviate from expectations drawn from tidal hydrodynamics \citep{Lightbody:2006vm}.  

Contemporary understanding of the fluid dynamics within vegetation canopies highlights the importance of coherent vortex structures resulting from interactions between canopy elements and the atmospheric boundary layer \citep{Belcher:2012bw,Nepf:2012ei}. In order to interpret our measurements in the context of these features of canopy turbulence, we make high-temporal-resolution measurements that allow for evaluation of the effects of shear-layer-induced billows and other short-time-scale motions.
%evcp

\section{Materials and Methods}
\subsection{Field Site}

Our study site was a 3-ha marsh located on Twitchell Island in the Sacramento-San Joaquin Delta of Northern California, USA (latitude: 38.1072$^\circ$ N; longitude: 121.6463$^\circ$ W), approximately 100 km inland from where the Sacramento-San Joaquin river system terminates in San Francisco Bay. %evcp
Late-nineteenth and early-twentieth century draining of the inland marshes in this region led to oxidation of the peat soil and many areas, including our study site, have subsided meters below sea level \citep{Rojstaczer:1995cy}. Our study site was used for farming until its restoration by the California Department of Water Resources (CA DWR) and the United States Geological Survey (USGS) in 1997. Small berms were built to isolate the 3-ha parcel and \textit{Schoenoplectus acutus} (Tule) shoots were planted across approximately 3\% of the area. \textit{Typha spp.} (cattail) also colonized significant areas via wind-blown propagules \citep{Miller:2008th} and by 2006 a canopy of \textit{Typha spp.} and \textit{S. acutus} covered 95\% of the marsh with an average density of 60 stems per square meter \citep{Miller:2009ju}. This density is on the lower end of naturally-occurring vegetation density, based on a literature survey of wetland plants similar to Tule and Typha \citep{CPthesis}.  Expressing the density as frontal area per area of wetland footprint, we calculate a nondimensional density of 0.15, which is about half that of the ``sparse'' case studied by \citep{Bane}.

The restored marsh is too subsided to receive water directly from the neighboring San Joaquin River, thus the water supply is delivered in a controlled manner by siphoning from the river to inlets located on the southern edge of the marsh. The water depth in the marsh ($h_w=0.25 m$) is maintained by controlling the height of an outflow weir located on the northern edge of the marsh.

\subsection{Ambient Conditions}

Data were recorded between 11:00 and 16:00 on 9 April 2013. The day was selected because of its unusually strong winds, which were consistent over the study period. The atmospheric stability during our experiment was between neutral and slightly unstable (nondimensional atmospheric stability parameter ranged from -0.01 to 0.00, computed from the 5-meter tower discussed below).
%evcp
Days before the experiment, flow from the two freshwater inlets were shut off, allowing the water level in the marsh to equilibrate with the outlet weir. Thus there should be negligible advection driven by flow into or out of the marsh during our measurement period. 

We deployed measurement equipment at the end of a wooden walkway extending into the interior of the marsh from the southern edge. During the duration of the measurements, the prevailing winds blew from north to south immediately over a 50 m stretch of marsh unobstructed by trees or other landscape features whose wake signatures could interfere with our measurements. 

The plant canopy surrounding our measurement site was comprised primarily of desiccated \textit{S. acutus} and \textit{Typha spp.} culms that, on average, reached a height of $h_c=1.75\pm0.25$ meters above the water surface, making the nondimensional water depth $h_w/h_c=0.125$.  The portions of the culms near the water surface were not observed to move with the wind and direct stirring of the water column by the vegetation was thus unlikely to have contributed significantly to surface water flow.  Similarly, observations by eye indicated no fluctuations in the water surface elevation, suggesting that progressive waves were not present during the study period.

The surface of the water in the area near the boardwalk was covered with a network of common duckweed (\textit{Lemna minor}) that was removed prior to our measurements. Surfactants such as biogenic lipids inhibit air-water momentum exchange and were likely present in this highly productive wetland as they are in most biologically active aquatic zones \citep{Frew:1990jt,McKenna:ul}.

\subsection{Measurement Equipment}

We simultaneously measured wind velocity below the top of the canopy and surface water velocity. Wind velocities were measured with a Young Model 81000 triaxial ultrasonic anemometer positioned with its measurement volume $1.35\pm0.01$ meters above the water surface. The anemometer was fastened atop a frame constructed from t-slotted aluminum extrusions. Water velocity was measured using the Volumetric Particle Imager (VoPI), an optical tool designed in-house to measure fluid velocities outdoor in hard-to-reach environments. Details of the VoPI's inner workings and performance can be found in \citep{Tse:2013co}, but in brief it is a video recording system designed to record 3D particle tracks within a cubic-centimeter sample volume. The VoPI was fastened to the bottom of the aluminum frame via a 2-axis stage mount that allowed translation along the vertical axis and one horizontal axis. The traversing horizontal axis was defined as the \textit{x}-axis, which we will refer to as the streamwise direction because it is closely aligned with the mean wind direction. The positive \textit{x} direction pointed toward $158^\circ$ magnetic N and the positive \textit{z}-axis pointed upward. The alignment of the two devices was estimated to have been within $\pm10$ degrees of one another. A portable marine battery provided power to both the sonic anemometer and the VoPI, with both its camera (Allied Vision Technologies Prosilica GC660) and its light source (Gradient Lens/Ushio Inc. Luxxor 24W LS) receiving 12 volts DC each. Data acquisition and storage for both devices were performed on a single PC, allowing data collection to be synchronized to a single internal clock and providing consistent timestamps for post-processing. For this study, the sonic anemometer collected data at 20 Hz, while the VoPI collected data at 80 Hz. The faster sample rate of the VoPI allowed for improved data resolution by track-smoothing interpolation.

Collaborating scientists from UC Berkeley had previously installed long-term meteorological recording equipment atop a 5-meter tower at the downwind end of the marsh (150 m to the east).  We conducted our within-canopy measurements at the closest accessible location upwind of this tower where the canopy was fully intact.  On the 5-meter tower, a sonic anemometer (Wind Master 1590, Gill Instruments, Lymington, United Kingdom) recorded wind velocity at 10 Hz from a height of 4.7 m. We used this dataset to calculate the statistics of the wind velocity above the vegetation canopy. The canopy structure and other landscape features were qualitatively the same in both wind-measurement locations; thus, we consider it appropriate to use the above-canopy wind measurements to inform our understanding of in-canopy flow behavior.

\subsection{Data Collection and Processing}

Four 10-minute measurements were collected in total. The first of four measurements, which we call Set A, was made with the distal tip of the VoPI positioned approximately 2 mm below the surface of the water, which was the shallowest depth attainable. The second measurement (Set B) was made with the VoPI positioned 40 mm below the surface at the same horizontal location as Set A. The third measurement (Set C) was made at a depth of 40 mm below the surface, but the VoPI was translated approximately 100 mm north (negative $x$ direction) relative to where the first two measurements were made. The fourth measurement (Set D) was made with the VoPI at the water surface at this new horizontal position. Sets C and D were taken to evaluate horizontal spatial variability.  Given the small size of the VoPI sample volume, there was no blockage by vegetation.

Images recorded by the VoPI were processed using the algorithms and procedures described in Tse and Variano \citep{Tse:2013co}, giving three-dimensional (3D) positions of suspended particles imaged in each frame. Particle trajectories could then be computed by linking corresponding particles across successive images. Particle-following, or Lagrangian, velocities from each trajectory were computed by numerically differentiating points along particle paths. Lastly, the average of all Lagrangian velocities at each time-step were transformed into the Eulerian velocity time series on which we performed our analysis.  Such averaging is justified by the observation that the variability in velocity across multiple particles at any given time is much less than the variability in velocity at one location over time.  An example movie showing the VoPI images is included as supplementary material.

\section{Results}

Velocity moment statistics computed from each 10-minute velocity record set are displayed on Tables \ref{tab:canopy} and \ref{tab:water}. The steadiness of the wind conditions was evident in the fact that velocity statistics converged well within a 10-minute window. The fourth statistical moment (kurtosis) on the water side has not converged completely, which is to be expected for a distributions with such large kurtosis values. %evcp
From Tables \ref{tab:canopy} and \ref{tab:water}, we see that while mean in-canopy wind velocities did not change appreciably from set to set, differences in the mean streamwise water velocities between the initial location (Sets A and B) and the second location (Sets C and D) indicate a degree of spatial variability. Although mean water velocities differed between the locations, velocity standard deviations and velocity power spectra (discussed later) remained consistent across all sets. The consistency of the higher moments and the spectral content convinced us that the spatial variability did not affect the fluctuating velocities, which we will later argue are more dynamically important for this system than mean values.

\begin{figure}[h] 
\centering
\includegraphics[scale=1.5]{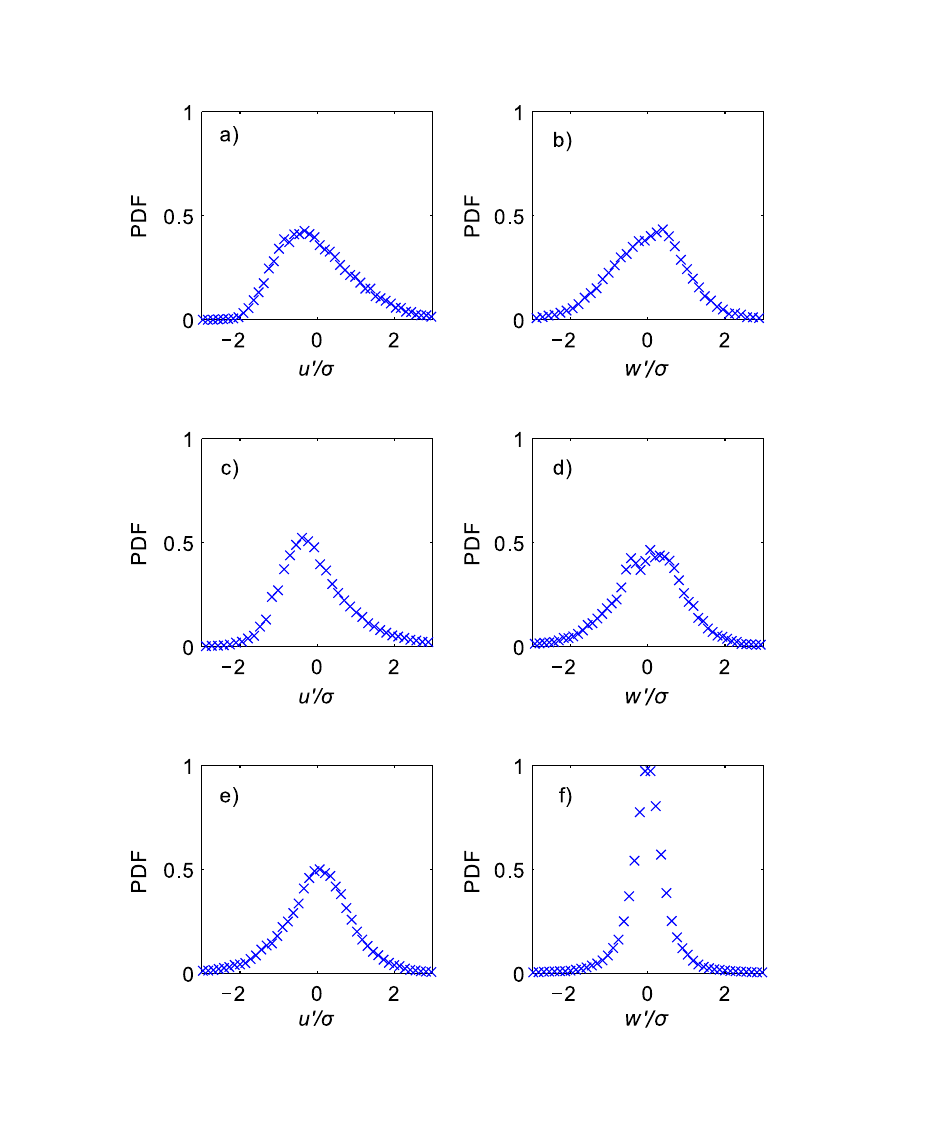}
\caption{Empirical probability density functions (PDF) of streamwise (\textit{u}) and vertical (\textit{w}) velocity fluctuations normalized by corresponding velocity standard deviations ($\sigma$). (a,b) Above-canopy wind. (c,d) In-canopy wind. (e,f) Surface water.}
\label{fig:pdf}
\end{figure}

\begin{figure}[h] 
\centering
\includegraphics[scale=0.2]{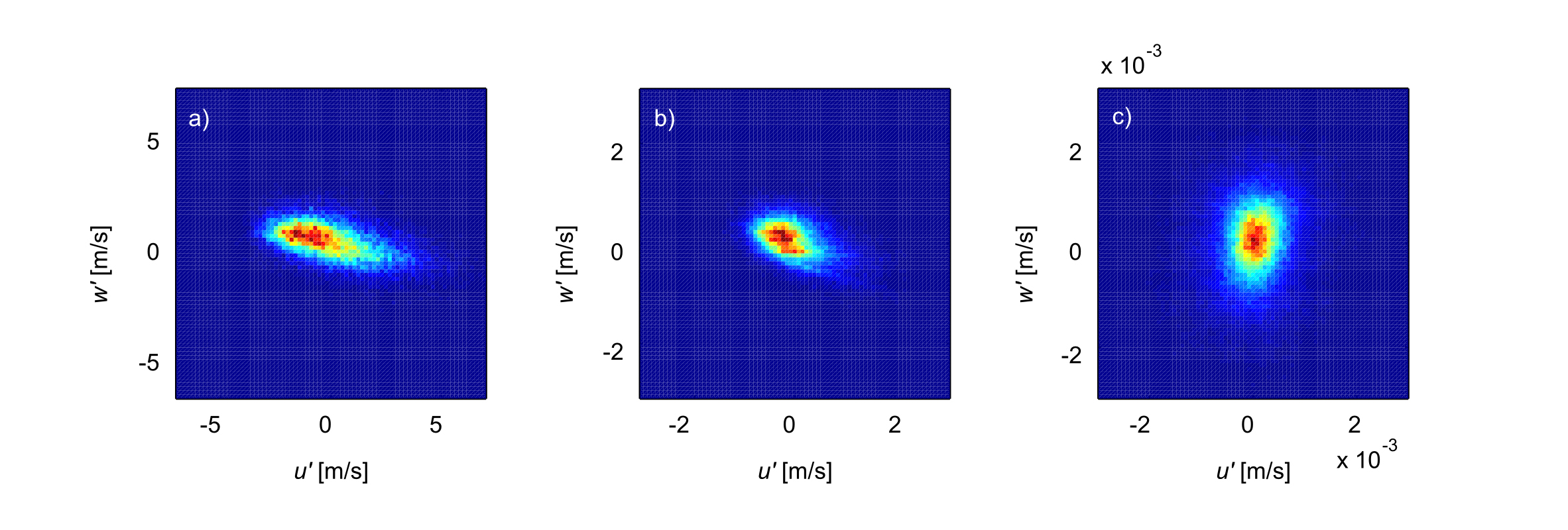}
\caption{Quadrant plot of the relative density of the $u'$-$w'$ correlation. (a) Above-canopy wind. (b) In-canopy wind. (c) Surface water. Data for (b,c) are combined from Sets A-D for in-canopy and surface water respectively. Colors correspond to the relative frequency of velocity correlations, where red is frequent occurrence and blue is near zero occurrence.}
%evcp
\label{fig:quad}
\end{figure}

Figure \ref{fig:pdf} shows the total distribution (with sets A-D combined) of the temporally fluctuating component of air and water velocities measured at each height. The distributions of wind measurements above and within the canopy exhibit noticeable skewness, with the streamwise component skewed positively and vertical component skewed negatively. The water-side velocity, however, was more symmetrically distributed around its mean relative to the air-side velocities. Figure \ref{fig:quad} shows quandrant plots displaying the density of the combined $u'$-$w'$ correlation measured at the three heights. Streamwise and vertical wind motions were strongly negatively correlated with each other, at locations in and above the canopy. These negative correlations indicate production of turbulent kinetic energy and the vertical transport of horizontal momentum.  Such patterns are typically caused by coherent vortex structures %evcp
that inject high-momentum fluid downward (``sweeps") and lift low-momentum fluid upwards (``ejections") \citep{pope2000turbulent}. The $u'$-$w'$ correlation was much weaker in the water, suggesting a lack of coherent vortex structures in the vertical-although some of the decorrelation must be attributed to measurement noise inherent to the VoPI itself \citep{Tse:2013co}. This measurement noise (which is most evident at very low velocities) is random and is strongest in the vertical component due to anisotropy in the stereoscopic reconstruction.

\begin{figure}[h] 
\centering
\includegraphics[scale=1]{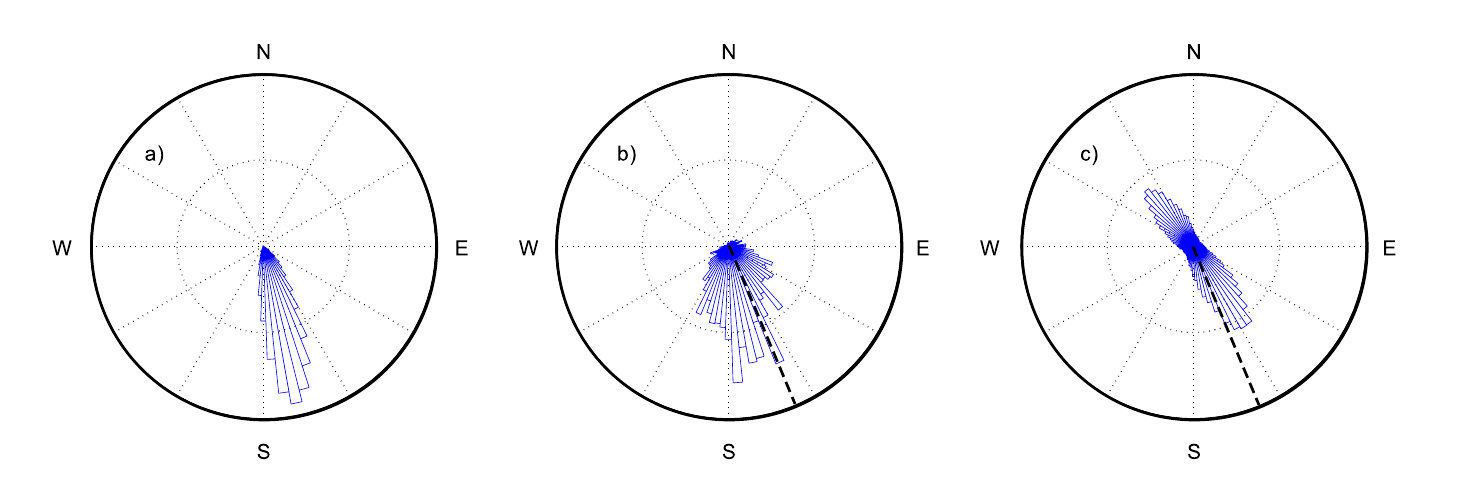}
\caption{Empirical probability density of horizontal velocity headings of both wind and water in polar coordinates. (a) Above-canopy wind. (b) In-canopy wind. (c) Surface water. Data for (b,c) are combined from Sets A-D for in-canopy and surface water respectively. Dashed line indicates the orientation of the streamwise axis-coordinate (\textit{x}-axis) with respect to the cardinal directions, chosen to maximize U and minimize V.  The total areas of the blue bars are equivalent.}
\label{fig:rose}
\end{figure}

Figure \ref{fig:rose} shows distributions of horizontal wind and water velocity bearings in polar form. Horizontal water velocities were nearly aligned with the horizontal wind velocities, which was consistent with our assumption that wind was the dominant source of momentum into the wetland waters during the experiment.  The slight offset of approximately 10$^{\circ}$ between wind and water can be attributed to the instrument misalignment during field deployment discussed in Section 2. Frequent occurrence of flow reversal on the water side was a prominent feature that contrasted with the unidirectional flow observed in the air (in Section 4, we argue that this oscillation is a result of the interaction between the water surface and packets of vortices generated in an upper-canopy shear layer). 

We examined the spectral content of wind and water velocity by computing velocity spectra from the time series data using two different methods. Some VoPI records contained blocks of missing data, so the spectra for the water-side velocity was computed via the Fourier transform of the autocorrelation function. The air-side data did not have gaps and its spectra were directly computed from the time-series. For both methods, we reduce noise by computing spectra over 30, 60 and 90 second subsets of each 10-minute timeseries, and then averaging.  The results were not sensitive to the subset length.  

\begin{figure}[h] 
\centering
\includegraphics[scale=1.2]{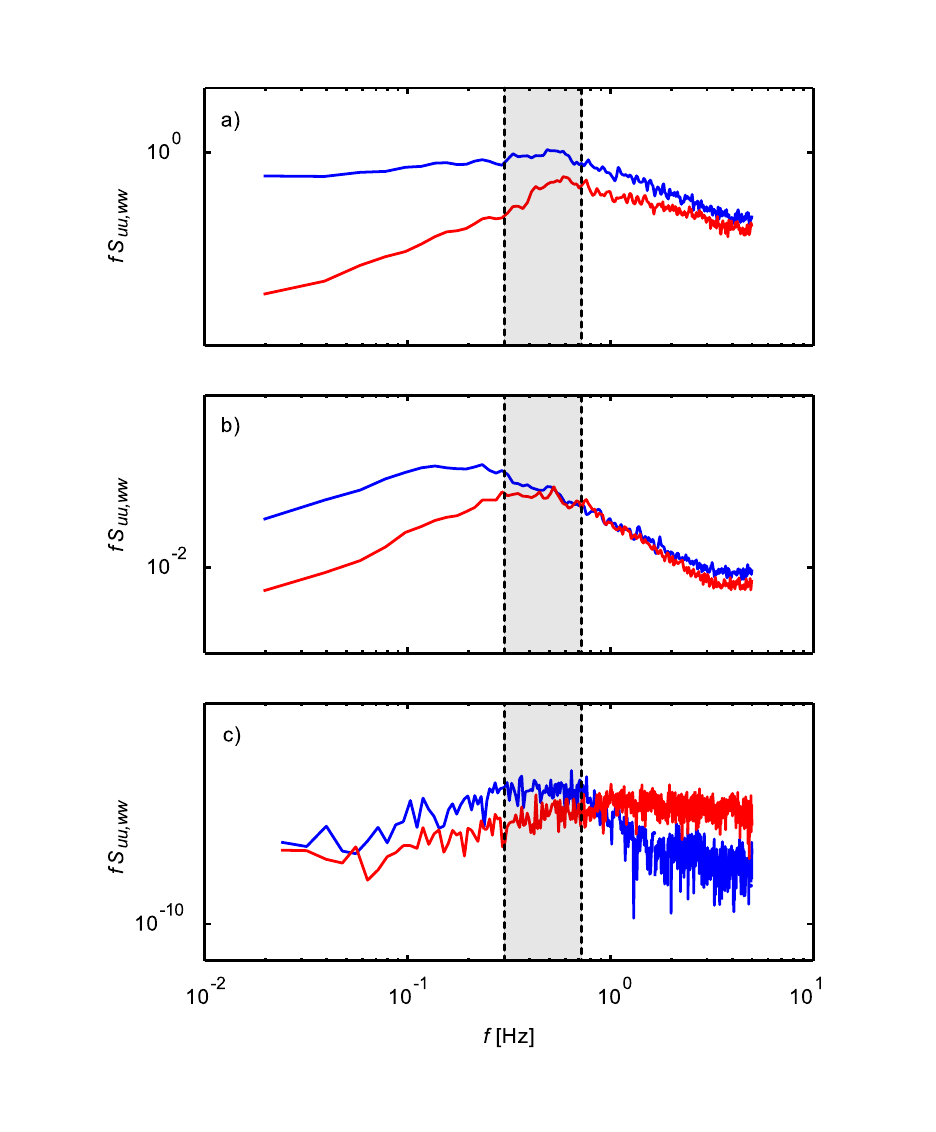}
\caption{Velocity spectra for the streamwise ($fS_{uu}$, blue) and vertical ($fS_{ww}$, red) fluctuating velocities, compensated by the frequency so that units are $m^2/s^2$. %evcp
(a) Above-canopy wind. (b) In-canopy wind. (c) Surface water. Data for (b,c) are combined from Sets A-D for in-canopy and surface water respectively. Shaded region on each panel indicates an approximate location of the broad peaks seen in some of these spectra.} %% is bounded by 0.3 and 0.7 Hz, which is the frequency range corresponding to the timescales of active coherent structures predicted by the mixing-layer analogy and dimensional scaling. In addition to being within this frequency range, the spectral peaks of the vertical velocity in (a,b) are well aligned with the spectral peak of the streamwise water velocity.}
\label{fig:spectra}
\end{figure}

Figure \ref{fig:spectra} shows the suite of velocity spectra computed for the three locations.  The spectra typically show peaks, which fall roughly in the range 0.3--0.7 Hz (except for the streamwise velocity peak in figure \ref{fig:spectra}b).  They are not extremely sharp peaks, and there is noise because of the short velocity timeseries, but the pattern is visible nonetheless.  In section 4 we propose a possible cause of this spectral peak.  It is unlikely that the peak is caused by waves, as no fluctuations in water surface elevation were observed during the measurement period.

Cross-correlations between the velocity fluctuations in-canopy and water-side were statistically identical to zero. In fact, no correlation peak emerged even when delays were set between +20 seconds and -20 seconds. This loss of correlation across the air-water interface is in line with the observations of Shaw et al. \citep{Shaw:1995wd}, who show that velocity correlation between two stationary points within a canopy rapidly decreases with increasing separation distance, especially in the horizontal direction. %evcp

\section{Discussion}
\subsection{Canopy Turbulence}

\begin{figure}[h] 
\centering
\includegraphics[scale=1.2]{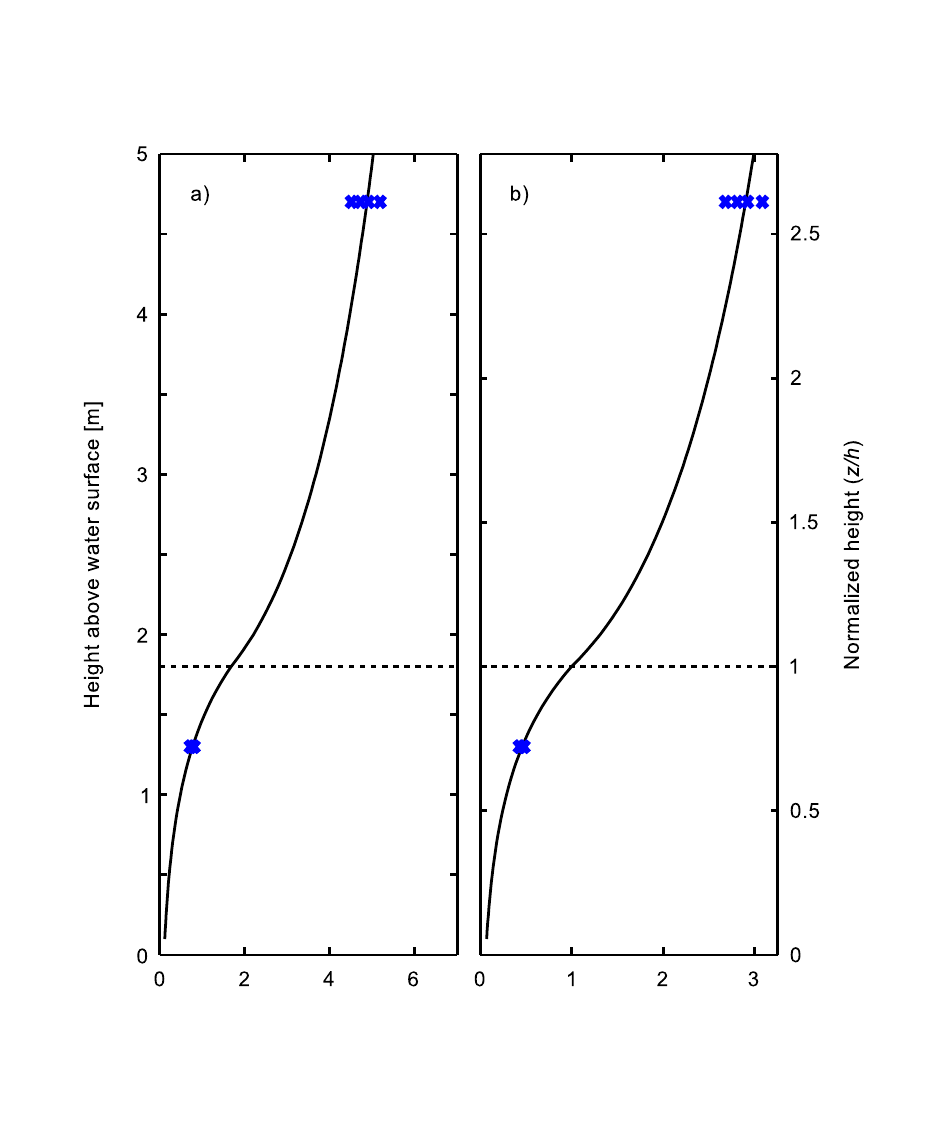}
\caption{Vertical profile of streamwise velocity computed from Equations \ref{eq:Uztop} and \ref{eq:Uzbot}. (a) Non-normalized profile. (b) Normalized profile: vertical position (\textit{z}) normalized by canopy height (\textit{h}) and velocity (\textit{U}) is normalized by the estimated velocity at canopy height ($U(h)$). Broken lines indicate the top of the canopy ($h=1.75$ m). Blue markers are mean streamwise wind velocities from in-canopy data Sets A-D (Table \ref{tab:canopy}) and from above-canopy data collected at the 5-meter tower measurements (Table \ref{tab:water}).}
\label{fig:profile}
\end{figure}

The flow structure within and just above vegetation canopies is well described using an analytical framework based on the plane mixing-layer, one of the canonical shear flows in turbulence research \citep{pope2000turbulent}. This ``mixing-layer analogy" provides an analytical description of velocity measurements and has been corroborated across a large range of canopy species, heights, densities, and even for fully submerged plant canopies \citep{Raupach:1996cj,Finnigan:2000vi,Ghisalberti:2002do}. The central focus of the mixing-layer framework is its consideration of canopy-scale coherent eddy structures. It states that aerodynamic drag exerted by canopy elements attenuates mean velocities exponentially with depth into the canopy, creating a strong inflection in the velocity profile at the canopy top (see Figure \ref{fig:profile}). Furthermore, instabilities arising from this profile inflection are mechanistically equivalent to those in classic mixing-layers; thus we expect the wind field around the canopy top to resemble coherent Kelvin-Helmholtz billows. These wind vortices are advected downstream as they form and can penetrate the entire depth of the vegetation canopy, depending on the strength of winds and the density of plant elements \citep{Finnigan:2000vi}.  These structures also evolve further through inertial instabilities and by interaction with plant stems, leading to a complex set of 3-dimensional coherent structures \citep{Pancham}. %evcp

The mean streamwise separation between successive eddies ($\Lambda_{x}$), or streamwise periodicity, represents the length scale of the dominant coherent motions at the canopy scale. Coherent billows can be identified in our velocity measurements by relating $\Lambda_{x}$ to the peak frequency of the in-canopy velocity spectra seen in Figure \ref{fig:spectra}. A reasonable prediction for $\Lambda_{x}$ is made using the relationship: $\Lambda_{x}/\delta_{w}\approx3$--5 (\citep{Raupach:1996cj}), where $\delta_{w}$ (the vorticity thickness) can be approximated from $L_{s}$, a characteristic length scale of the velocity profile ($\delta_{w}\approx2L_s$). Thus, the following relationship between streamwise periodicity and the shear scale was used, $\Lambda_{x}\approx8L_s$ which has performed well across many types of vegetation canopies \citep{Raupach:1996cj,Finnigan:2000vi}. A basic shear length scale for canopy flows is taken as the time-averaged horizontal velocity at the canopy top ($U(h)$) divided by the shear ($\partial U/\partial z$) there:

\begin{equation} \label{eq:Ls}
L_s=\frac{U(h)}{\left.\frac{\partial U}{\partial z}\right|_{z=h}}
\end{equation}

To estimate $L_s$ for our data, we employed a simplified model to estimate $U(h)$. Using data from the sonic anemometer within the canopy ($z/h=0.72$) and from the nearby meteorological tower ($z/h=2.6$) as reference points, we modeled the mean velocity above the canopy using:

\begin{equation} \label{eq:Uztop}
U(z)=\frac{u^\ast}{\kappa}\ln\left(\frac{z-z_0}{d}\right)
\end{equation}

Where $\kappa$ is the von Karman constant ($\kappa=0.4$). Computed with the fluctuating part of the streamwise ($u'$) and vertical components ($w'$) of wind velocities as $u^\ast=\langle u'w'\rangle ^{\frac{1}{2}}$, the friction velocity ($u^\ast=0.9\pm0.1$ m/s) was measured at the met tower ($z/h=2.6$) because there is less vertical variation in shear stress in a region far removed from the velocity inflection \citep{Finnigan:2000vi}. The values for the roughness height ($h_0$) and zero-plane displacement ($d$) were fit to the measurement data using $z_0/h\approx0.1$ and $d/h\approx0.65$ as initial reference values that were typical for tall crop canopies \citep{campbell1998introduction}. The best-fit curve in Figure \ref{fig:profile} was drawn using $z_0/h=0.13$ and $d/h=0.67$ m.  The log-law profile was matched to an exponential profile at an inflection point set exactly at the canopy height.  The exponential profile of mean in-canopy velocity was modeled following \citep{cionco} as: %evcp

\begin{equation} \label{eq:Uzbot}
U(z)=U(h)\exp\left\{\alpha\left(\frac{z}{h}-1\right)\right\}
\end{equation}

Where $U(h)$ was the mean velocity estimated at $z=h$ using Equation \ref{eq:Uztop}, and the leaf-area drag ($\alpha=2.8$) was estimated from previous measurements of crop canopies with similar foliage type and vertical distribution \citep{campbell1998introduction}. From these curves, we found the shear length scale to be $L_s=0.58$ m, with $U(h)=1.7$ m/s and $\partial U/\partial z=2.9$ s$^{-1}$ using Equation \ref{eq:Ls}. Using the equation $\Lambda_{x}\approx8L_s$, we arrived at an estimate of $\Lambda_{x}\approx 4.6$ m. The length scale $\Lambda_{x}$ is related to the peak frequency ($f_p$) on our in-canopy velocity spectra via:

\begin{equation} \label{eq:fp}
f_p=\frac{U_{adv}}{\Lambda_x}\approx\frac{1.8U(h)}{\Lambda_x}
\end{equation}

Where the expression for advection velocity ($U_{adv}$) was taken from Raupach et al. \citep{Raupach:1996cj}, who argued that the velocity at which these eddies are traveling through the canopies is underestimated by $U(h)$ and that the true velocity at which billows are advected through a canopy is closer to $U_{adv}\approx1.8U(h)$. Using Equation \ref{eq:fp}, we estimated the dominant frequency of the coherent motions to be $f_p\approx0.7$ Hz, a value consistent with the range of peak frequencies we measured in the canopy and in the water.  %Those peaks, seen in Figure \ref{fig:spectra}, fall roughly in the range 0.3--0.7 Hz, other than the streamwise velocity peak in figure \ref{fig:spectra}b.  They are not extremely sharp peaks, and there is noise because of the short velocity timeseries, but the pattern is visible nonetheless. 
One explanation for why the vertical fluctuating in-canopy velocity spectra more closely agrees with predictions from the mixing-layer analogy than the streamwise spectra does is because the length-scales of the vertical motions are more directly linked to the unique dynamics determined by the canopy geometry, while horizontal velocities are heavily influenced by large-scale atmospheric events that may not be involved in canopy-scale transport \citep{Raupach:1996cj}.

\subsection{Water-side Motions}

Wind-induced momentum flux across an air-water interface under wave-free conditions depends on the shear stress at the interface. In open water environments, i.e., lakes and open seas, where the atmospheric boundary layer above the interface is turbulent and takes on the classic logarithmic profile, scientists typically estimate shear stress using mean wind velocities measured at the height of 10 meters above sea level ($U_{10}$) \citep{thorpe2005turbulent}. However, given our data and the mixing-layer analogy, in a dense vegetation canopy the average free-stream velocity above the canopy poorly represents the dynamics of the airflow just above the air-water interface. Specifically, the periodic nature of water motions cannot be predicted by above-canopy winds except through the scaling arguments presented earlier in the discussion.

\begin{figure}[h] 
\centering
\includegraphics[scale=1.2]{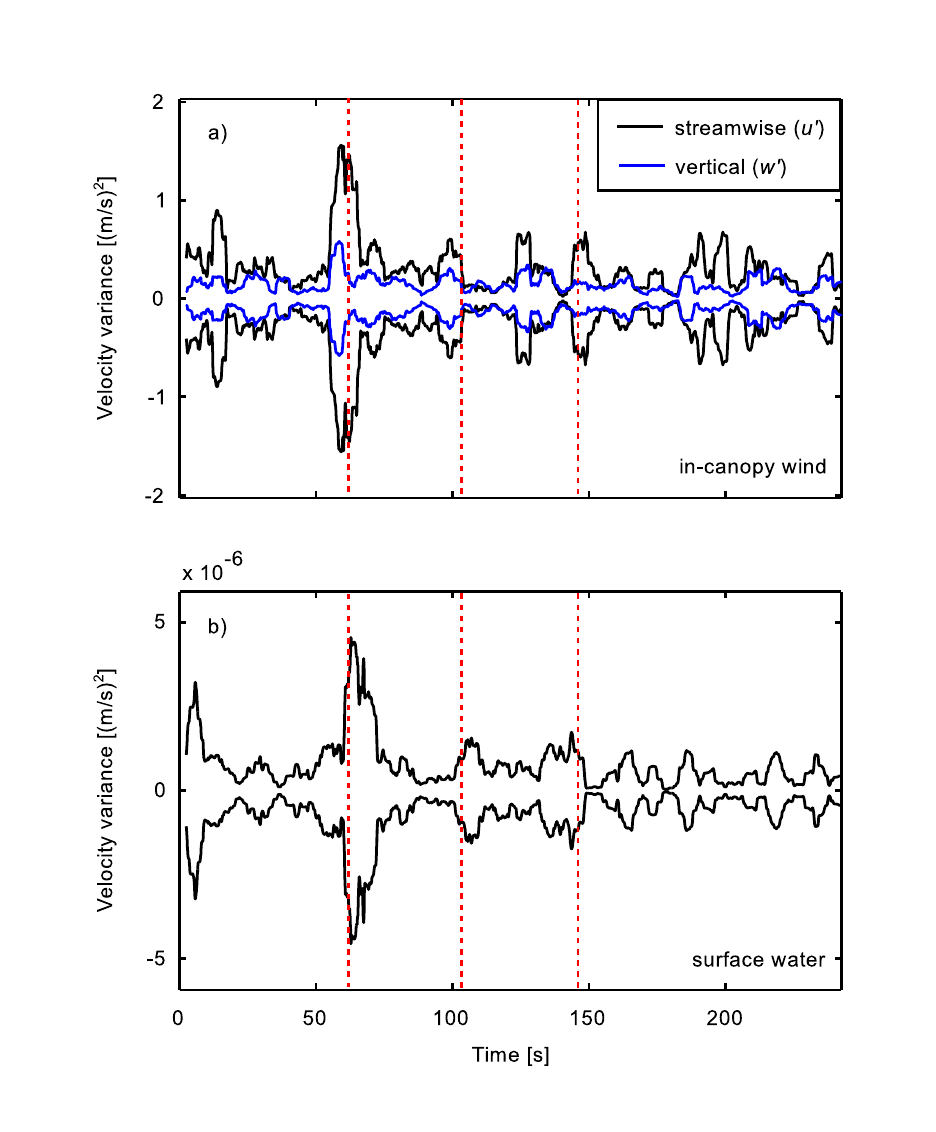}
\caption{Velocity variance computed with a 5-second moving window ($N=100$) on a 5-minute segment from measurement Set C. (a) Streamwise ($u'$) and vertical ($w'$) in-canopy wind velocity variances. (b) Streamwise ($u'$) surface water velocity variance. Dotted lines indicate several instances of high correlation between the variance measurements of in-canopy and surface water velocities.}
\label{fig:var1}
\end{figure}

\begin{figure}[h] 
\centering
\includegraphics[scale=1]{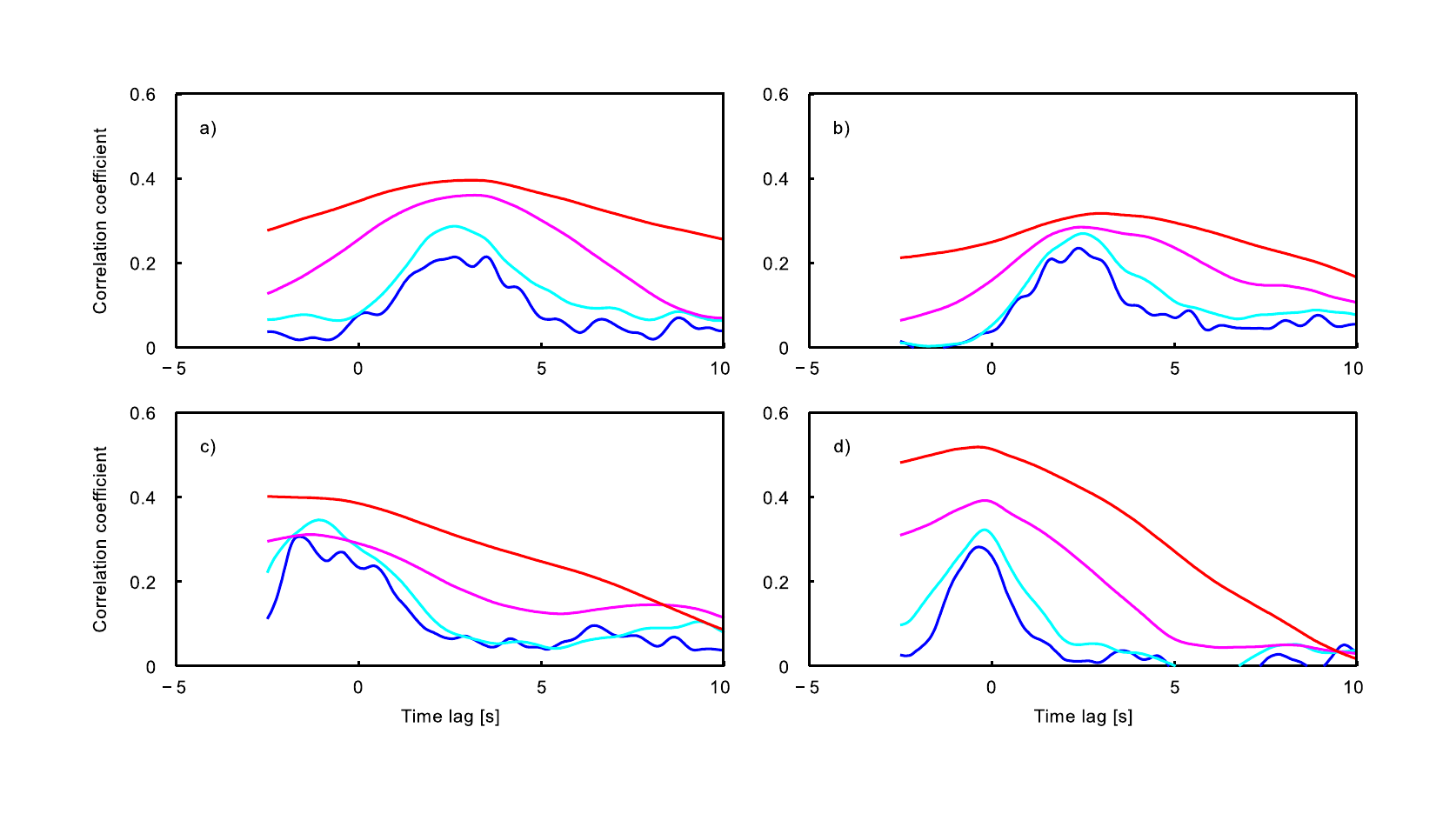}
\caption{Correlation coefficient of running velocity variance values as a function of time lag. Panels (a-d) correspond to the measurement Sets (A-D) respectively. Each curve corresponds to different window sizes: blue corresponds to a 1 second window ($N=20$), cyan to a 2 second window ($N=40$), magenta to a 5 second window ($N=100$), red to a 10 second window ($N=200$).}
\label{fig:var2}
\end{figure}

Two connections between water velocities and in-canopy wind velocities strongly support an interfacial momentum transfer model based on canopy turbulence. First, the spectral peak in the streamwise velocity of the water is aligned with the spectral peaks in the in-canopy spectra (Figure \ref{fig:spectra} panels a--c). Given the cross-sectional dimension of the aluminum frames and the average in-canopy wind speeds, potentially obscurant velocity perturbations caused by vortex shedding off of the measurement equipment should appear at frequencies of roughly 30 Hz, which is much higher than those seen in Figure \ref{fig:spectra}. %evcp
Absent other drivers of momentum exchange, the alignment in the spectral peaks of both air and water suggest Kelvin-Helmholtz billows were directly responsible for the motions measured in the water.  The second connection we observe between water velocities and in-canopy wind is a strong correlation in the temporal pattern of velocity variance. This is related to the observation that water velocities showed episodic bursts of high-amplitude fluctuations. Such behavior is seen in Figure \ref{fig:var1}, which is a time-series of velocity variance computed over a running 5-second window. The results show several instances of high-energy events (high velocity variance) that appear concurrently in both air and water. The similarities of both the onset and duration of these events suggest they may be triggered by the same billow events. We quantify the co-occurrence of high-variance events in air and water by calculating the cross-correlation of the variance time-series (Figure \ref{fig:var2}). %evcp
The velocity variance was well correlated across the air-water interface regardless of the size of the averaging window, which we varied from 1 to 10 seconds. The strength of the correlation is quite remarkable, given that velocity time-series rapidly decorrelate with distance inside a plant canopy. In contrast, we find that the velocity variance remains well correlated across 1.35 m of height and the air-water interface. The difference in these two correlation behaviors suggests that billows transport bursts of high-energy motion through the plant canopy and into the water column, but that the detailed kinematic expression of the billow changes rapidly through space as it interacts with the vegetation canopy.

From Figure \ref{fig:var2}, we can also see that the correlation peaks for Sets A and B occur at bigger lag times than those from Sets C and D, suggesting that lags in response time depend more on horizontal location than on water depth. While we do not have a compelling explanation for this observation, we do observe a similar dichotomy in the mean velocities between the first (A and B) and second (C and D) horizontal locations. This dependence on horizontal position may be a consequence of the interaction of wind with the heterogeneous distribution of plant elements.  The fact that horizontal variation adds nuance to the atmosphere-water coupling, but does not affect the magnitude of velocity variance correlations, lends further support to our hypothesis of billow-driven momentum transport across the air-water interface.

\subsection{Stirring in the water column}
Poindexter and Variano's laboratory measurements of interfacial gas transfer velocities showed that under high wind conditions (i.e., in-canopy windspeed of 1 m/s), wind-driven stirring caused air-water exchanges similar in magnitude to typical nighttime thermal convection \citep{Poindexter:2013kg} . We directly confirmed their result here by showing that the velocity scale (which was most appropriately the velocity variance in our case) in the surface waters of a canopied wetland on a very windy day was similar to convective velocity scales associated with thermal convection driven by a water column heat loss rate of $q=-200$ W/m$^2$. In such a case, the convective velocity scale $w^\ast=(Bh)^{\frac{1}{3}}$, where $B$ is the buoyancy production and $h$ the water depth, is on the order of $1\times10^{-3}$ ms$^{-1}$, the same order of magnitude as the velocity variance observed here (see Table \ref{tab:water}).  Our data also indicate, however, that the water-column motions responsible for stirring look very different than turbulence.

The Lagrangian particle trajectories from the VoPI indicate that the water flow did not exhibit any of the characteristic motions of turbulence, such as multiple timescales and curved pathlines. Moreover, the dominant motions were also confined to a narrow frequency range in contrast with the wide range of time scales characteristic of fully developed turbulence.  

The trajectories also did not show evidence of progressive wave motion, which supports the observed lack of water surface fluctuations.  Interestingly, the video measurements did show particles undergoing back-and-forth motions, but there was usually only one cycle, not a continuous oscillation.  That is, the motions appear more like solitary waves than a wave train. We consider that this motion is a response to billows and related coherent structures, rather than progressive waves.

Although the energy flux into the water was insufficient to produce turbulence in this case, the flow patterns we measured were nevertheless capable of mixing and transport. For instance, cyclic water motion will affect the wake region around stems, where high surface divergence causes a local peak in scalar transport across the air-water interface \citep{Turney:2005tr,Poindexter:2013kg}.

Additional evidence that the wind-driven velocity fluctuations contributed to mixing in the water column can be seen by considering Stokes' second problem, which analytically describes the steady-state motion of an infinite body of fluid driven solely by an infinitely expansive flat surface at one boundary executing sinusoidal oscillations parallel to that interface. The solution to the governing equation shows that the extent of the momentum transfer is confined within a thin layer of depth $\delta\approx4\sqrt{\nu/\omega}$ adjacent to the interface, where $\nu$ is the kinematic viscosity and $\omega$ is the oscillation frequency of the interface \citep{kundu2012fluid}. Given an interfacial forcing frequency of 0.7 Hz (the peak of the in-canopy wind spectrum), Stokes' solution predicts a surface layer penetration depth of only 2 mm. In contrast, water-side variance was coupled with air-side variances at least down to 4 cm. This thicker layer suggests that the water column motion caused by the billows facilitated momentum transport at a rate faster than by viscosity alone.  Using Stokes' solution, we can estimate an Eddy viscosity of 5 cm$^2$/s during the observation period.

Considering the above, we conclude that stirring near the water surface is (a) strong enough to influence biogeochemical budgets and (b) cause primarily by transfer of momentum from above the canopy, through the air in the canopy, and across the surface via shear.  That is, neither waving plants nor progressive waves on the surface are strong contributors to surface water motion at the study site under the study conditions.

\section{Conclusion}

Here, we present the first field 
%evcp
measurement of wind-driven momentum transfer across the air-water interface within a wetland with emergent vegetation. The study was conducted on a single day with interesting atmospheric conditions, i.e. strong winds and neutral atmospheric stability.  
%evcp
Simultaneous measurements of in-canopy wind velocities using a sonic anemometer and surface water velocities using a 3D particle-tracking camera revealed that canopy-scale eddies not only dominate the dynamics of airflow within the plant canopy but also the dynamics of the surface water beneath the canopy. The transport of momentum from the atmosphere to the water column was mediated by the dynamics of canopy turbulence, meaning that the conventional relationship relating the interfacial momentum flux to the average free-stream windspeed is an insufficient description when in the presence of a dense vegetation canopy. The mixing-layer analogy, which describes the airflow within and just above a terrestrial plant canopy as being characterized by billow structures borne from instabilities, can be used to predict the dominant scales of motion transferred across the air-water interface. Our measurements showed that under strong, persistent wind forcing, surface waters oscillated at a range of frequencies, but with a distinct peak at the same frequency the streamwise periodicity of the canopy-scale billows. These wind-driven water motions can carry momentum deeper into the water column than viscous transport alone, and augment scalar transport at the air-water interface.

%%% End of body of article:

%%%%%%%%%%%%%%%%%%%%%%%%%%%%%%%%
%% Optional Appendix goes here
%
% \appendix resets counters and redefines section heads
% but doesn't print anything.
% After typing \appendix
%
%\section{Here Is Appendix Title}
% will show
% Appendix A: Here Is Appendix Title
%
%%%%%%%%%%%%%%%%%%%%%%%%%%%%%%%%%%%%%%%%%%%%%%%%%%%%%%%%%%%%%%%%
%
% Optional Glossary or Notation section, goes here
%
%%%%%%%%%%%%%%
% Glossary is only allowed in Reviews of Geophysics
% \section*{Glossary}
% \paragraph{Term}
% Term Definition here
%
%%%%%%%%%%%%%%
% Notation -- End each entry with a period.
% \begin{notation}
% Term & definition.\\
% Second term & second definition.\\
% \end{notation}
%%%%%%%%%%%%%%%%%%%%%%%%%%%%%%%%%%%%%%%%%%%%%%%%%%%%%%%%%%%%%%%%
%
%  ACKNOWLEDGMENTS

\begin{acknowledgments}
Support from the UC Berkeley department of Civil and Environmental Engineering allowed IT and CP to conduct these measurements, thereby overlapping their PhD dissertation projects.  Further analyses by CP were supported by California State University, Sacramento.  Two anonymous reviewers provided comments that improved the manuscript and analysis.  Data is available upon request from variano@berkeley.edu.
\end{acknowledgments}
%evcp

%% ------------------------------------------------------------------------ %%
%%  REFERENCE LIST AND TEXT CITATIONS
%
% Either type in your references using
% \begin{thebibliography}{}
% \bibitem{}
% Text
% \end{thebibliography}
%
% Or,
%
% If you use BiBTeX for your references, please use the agufull08.bst file (available at % ftp://ftp.agu.org/journals/latex/journals/Manuscript-Preparation/) to produce your .bbl
% file and copy the contents into your paper here.
%
% Follow these steps:
% 1. Run LaTeX on your LaTeX file.
%
% 2. Make sure the bibliography style appears as \bibliographystyle{agufull08}. Run BiBTeX on your LaTeX
% file.
%
% 3. Open the new .bbl file containing the reference list and
%   copy all the contents into your LaTeX file here.
%
% 4. Comment out the old \bibliographystyle and \bibliography commands.
%
% 5. Run LaTeX on your new file before submitting.
%
% AGU does not want a .bib or a .bbl file. Please copy in the contents of your .bbl file here.

%\bibliographystyle{agufull08}
%\bibliography{delta_wrr3}{}

%
% Please use ONLY \citet and \citep for reference citations.
% DO NOT use other cite commands (e.g., \cite, \citeyear, \nocite, \citealp, etc.).

%% ------------------------------------------------------------------------ %%
%
%  END ARTICLE
%
%% ------------------------------------------------------------------------ %%
\end{article}
%
%
%% Enter Figures and Tables here:
%
% DO NOT USE \psfrag or \subfigure commands.
%
% Figure captions go below the figure.
% Table titles go above tables; all other caption information
%  should be placed in footnotes below the table.
%
%Figures
\begin{figure}[h] 
\centering
\includegraphics[scale=0.8]{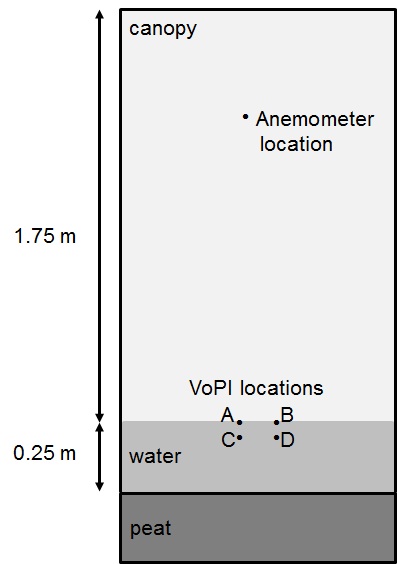}
\caption{Schematic diagram of key site features and measurement locations.  Not shown is a second anemometer 4.5 m above the water surface and 150 m downwind.}
\label{fig:schematic}
\end{figure}

%Tables
\noindent
\begin{table}[h]
\begin{tabularx}{\textwidth}{l | X X X | X X X | X X X | X X X }
\hline
	& \multicolumn{3}{c}{Mean [m/s]} & \multicolumn{3}{c}{Std. Dev. [m/s]} & \multicolumn{3}{c}{Skewness [-]} & \multicolumn{3}{c}{Kurtosis [-]} \\
	& U & V & W & U & V & W & U & V & W & U & V & W \\
\hline
	Set A & 0.81 & -0.21 & 0.16 & 0.66 & 0.57 & 0.41 & 0.87 & -0.85 & -0.33 & 4.3 & 6.3 & 4.7 \\
	Set B & 0.72 & -0.13 & 0.22 & 0.65 & 0.53 & 0.40 & 1.10 & -0.67 & -0.53 & 5.7 & 4.8 & 4.6 \\
	Set C & 0.80 & -0.21 & 0.18 & 0.66 & 0.57 & 0.41 & 0.83 & -0.72 & -0.22 & 5.2 & 5.6 & 4.5 \\		
	Set D & 0.73 & -0.14 & 0.20 & 0.64 & 0.56 & 0.40 & 1.10 & -0.73 & -0.38 & 5.2 & 5.4 & 4.4 \\ 
\hline
\end{tabularx}
\caption{Velocity moments measured by sonic anemometer within vegetation canopy of each orthogonal component of wind velocities for Sets A-D. \label{tab:canopy}}
\end{table}

\noindent
\begin{table}[h]
\begin{tabularx}{\textwidth}{l | X X X | X X X | X X X | X X X }
\hline
	& \multicolumn{3}{c}{Mean [$10^{-3}$ m/s]} & \multicolumn{3}{c}{S.D. [$10^{-3}$ m/s]} & \multicolumn{3}{c}{Skewness [-]} & \multicolumn{3}{c}{Kurtosis [-]} \\
	& U & V & W & U & V & W & U & V & W & U & V & W \\
\hline
	Set A & 3.7 & 0.30 & 3.5 & 0.85 & 0.49 & 0.69 & -0.44 & -0.13 & 0.86 & 5.2 & 5.8 & 32 \\
	Set B & 1.1 & 0.19 & 4.6 & 0.93 & 0.50 & 1.0 & -0.13 & 0.08 & 2.1 & 4.0 & 10 & 84 \\
	Set C & -1.5 & -0.70 & 4.0 & 0.92 & 0.75 & 1.3 & -0.63 & 2.4 & 0.63 & 6.8 & 172 & 77 \\
	Set D & -0.53 & 0.14 & 2.6 & 0.98 & 0.82 & 1.7 & -0.74 & 0.82 & -2.9 & 12 & 16 & 143 \\	 
\hline
\end{tabularx}
\caption{Velocity moments measured by the VoPI inside the water column of each orthogonal component of water velocities for Sets A-D.  \label{tab:water}}
\end{table}

\end{document}